\documentclass[10pt,twocolumn,letterpaper]{article}

\usepackage{cvpr}              

\usepackage{graphicx}
\usepackage{amsmath}
\usepackage{amssymb}
\usepackage{booktabs}

\usepackage{amsmath}
\usepackage{algorithmic}
\usepackage{color}

\usepackage{multirow}

\usepackage{arydshln}
\usepackage{lipsum}
\usepackage{anyfontsize}
\usepackage{t1enc}

\usepackage[table]{xcolor}

\usepackage[pagebackref,breaklinks,colorlinks]{hyperref}

\usepackage{array}
\usepackage[capitalize]{cleveref}
\crefname{section}{Sec.}{Secs.}
\Crefname{section}{Section}{Sections}
\Crefname{table}{Table}{Tables}
\crefname{table}{Tab.}{Tabs.}

\def\cvprPaperID{2823} 
\def\confName{CVPR}
\def\confYear{2022}
\begin{document}
\title{Dual Adversarial Adaptation for\\ Cross-Device Real-World Image Super-Resolution}
\author{
    {
    Xiaoqian Xu$^1$ \space
    Pengxu Wei\thanks{Corresponding author: weipx3@mail.sysu.edu.cn}
    \space$^{1}$ \space
    Weikai Chen$^2$ \space
    Yang Liu$^1$ \space
    Mingzhi Mao$^1$ \space
    Liang Lin$^1$ \space
    Guanbin Li$^{1}$
    } \\
    {
    $^1$Sun Yat-sen University \space
    $^2$Tencent America
    }
}
\maketitle
\begin{abstract}
Due to the sophisticated imaging process, an identical scene captured by different cameras could exhibit distinct imaging patterns, introducing distinct proficiency among the super-resolution (SR) models trained on images from different devices.
In this paper, we investigate a novel and practical task coded \textit{cross-device SR}, which strives to adapt a real-world SR model trained on the paired images captured by one camera to low-resolution (LR) images captured by arbitrary target devices.
The proposed task is highly challenging due to the absence of paired data from various imaging devices. To address this issue, we propose an unsupervised domain adaptation mechanism for real-world SR, named \emph{Dual ADversarial Adaptation} (DADA), which only requires LR images in the target domain with available real paired data from a source camera. DADA employs the Domain-Invariant Attention (DIA) module to establish the basis of target model training even without HR supervision.
Furthermore, the dual framework of DADA facilitates an Inter-domain Adversarial Adaptation (InterAA) in \emph{one branch} for \emph{two LR input images} from two domains, and an Intra-domain Adversarial Adaptation (IntraAA) in \emph{two branches} for \emph{an LR input image}. InterAA and IntraAA together improve the model transferability from \textcolor{black}{the} source domain to the target. We empirically conduct experiments under six \emph{Real}$\rightarrow$\emph{Real} adaptation settings among three different cameras, and achieve superior performance compared with existing state-of-the-art approaches. We also evaluate the proposed DADA to address the adaptation to \textcolor{black}{the} video camera, which presents a promising research topic to promote the wide applications of real-world super-resolution.
Our source code is publicly available at
\href{https://github.com/lonelyhope/DADA.git}{https://github.com/lonelyhope/DADA}.
\end{abstract}

\section{Introduction}

\begin{figure}[t]
  \includegraphics[width=0.5\textwidth]{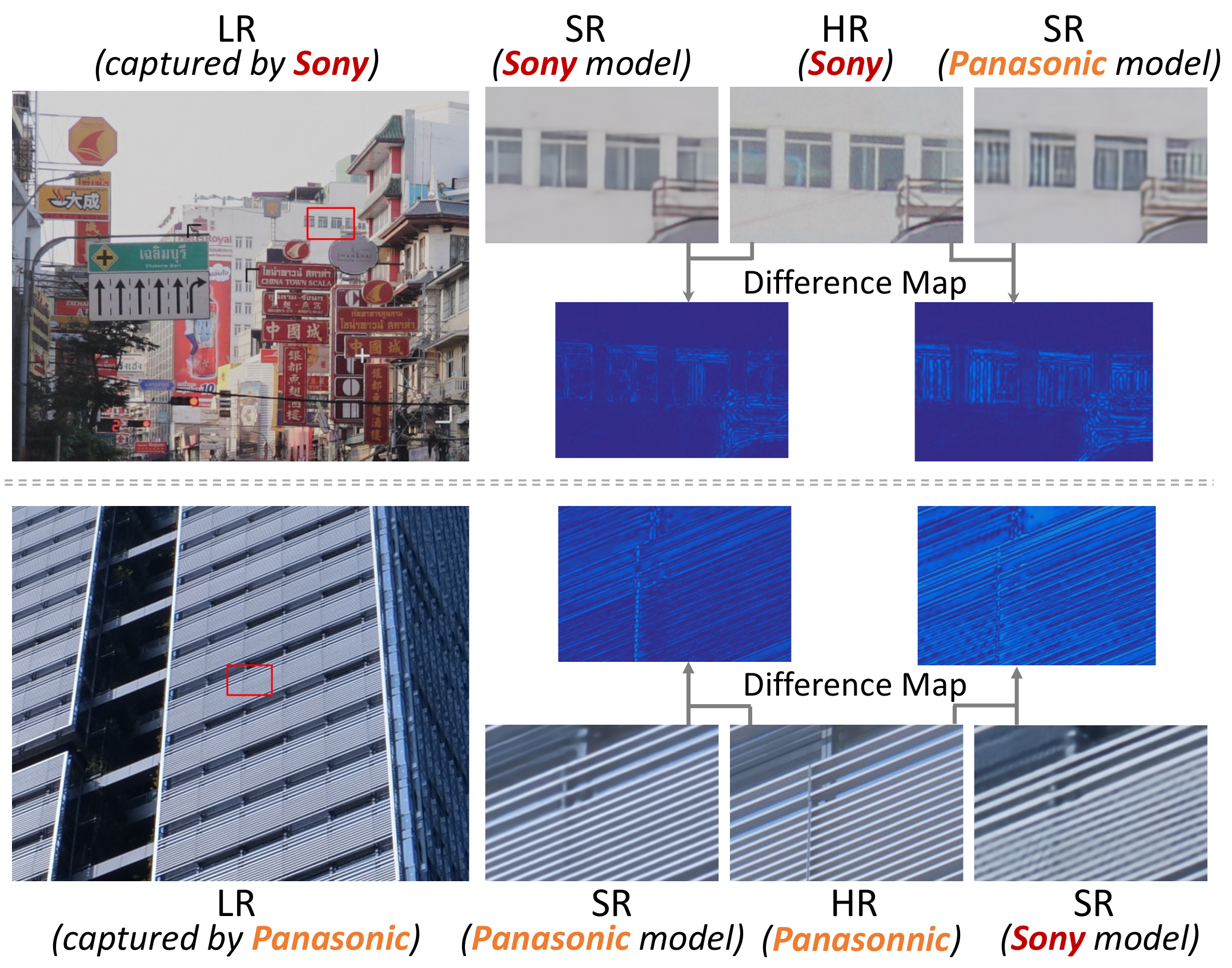}
  \vspace{-15pt}
  \caption{
  Comparison of real SR results across different trained models from different camera data in the DRealSR dataset~\cite{CDC}. \textcolor{black}{The Difference Map denotes the absolute difference between ground-truth HR and SR image (The brightness in the map reflects the magnitude of the difference
  ).}
  }
  \vspace{-10pt}
  \label{fig:diff}
\end{figure}
\begin{figure}[tp]
  \centering
  \includegraphics[width=.49\textwidth]{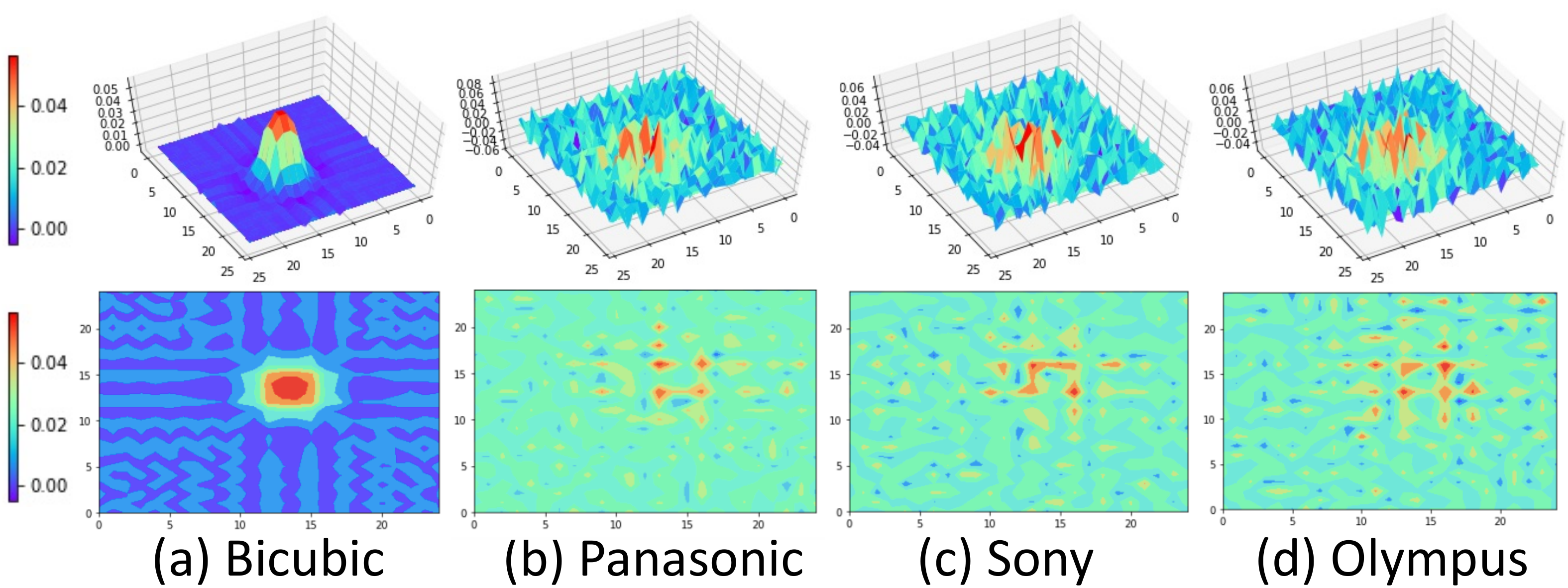}
  \vspace{-15pt}
  \caption{
  Degradation kernels for bicubic down-sampling and three different cameras in DRealSR \cite{CDC}.
  USRNet~\cite{USRnet} is employed to estimate those kernels by minimizing $\| (HR \otimes  k) \downarrow_s - LR \|$, where $\otimes$ means convolution,
  and $\downarrow_s$ means down-sampling for scaling factor $s$ by choosing the upper left item for every $s*s$ grid.
  The kernel size is set to $25*25$ and the scaling factor $s$ is 4.
  }
  \vspace{-10pt}
  \label{fig:kernel}
\end{figure}


Single image super-resolution (SISR), which super-resolves low-resolution images (LRs) and reconstructs their high-resolution counterparts (HRs), is a fundamental task in low-level computer vision.
The emergence of deep learning significantly contributes to the SR progress and SISR is usually cast as a supervised learning task with paired LR-HR images~\cite{Dong2014LearningAD,Ledig2017PhotoRealisticSI,Lim2017EnhancedDR,Wang2018ESRGANES,Zhang2018ImageSU}.
However, due to the difficulty of collecting LR-HR image pairs, typical \textcolor{black}{learning-based} methods learn to map synthetic low-resolution images to the original counterpart to achieve super-resolution reconstruction, which is constantly criticized with poor model generalization in practical scenarios. 
Accordingly, real SR has come up to explore the real image degradation and several real-world SR datasets collected by capturing paired LR-HR data via optical zoom of DSLR cameras emerged immediately, \emph{e.g.}, \emph{RealSR}~\cite{Cai2019TowardRS} and \emph{DRealSR}~\cite{CDC}. Considering the vulnerability of deep networks for quasi-imperceptible noises, robust real-world SR is further investigated to defense the adversarial attacks for practical applications~\cite{yue2021robust}.

\begin{figure}[t]
  \centering
  \includegraphics[width=.48\textwidth]{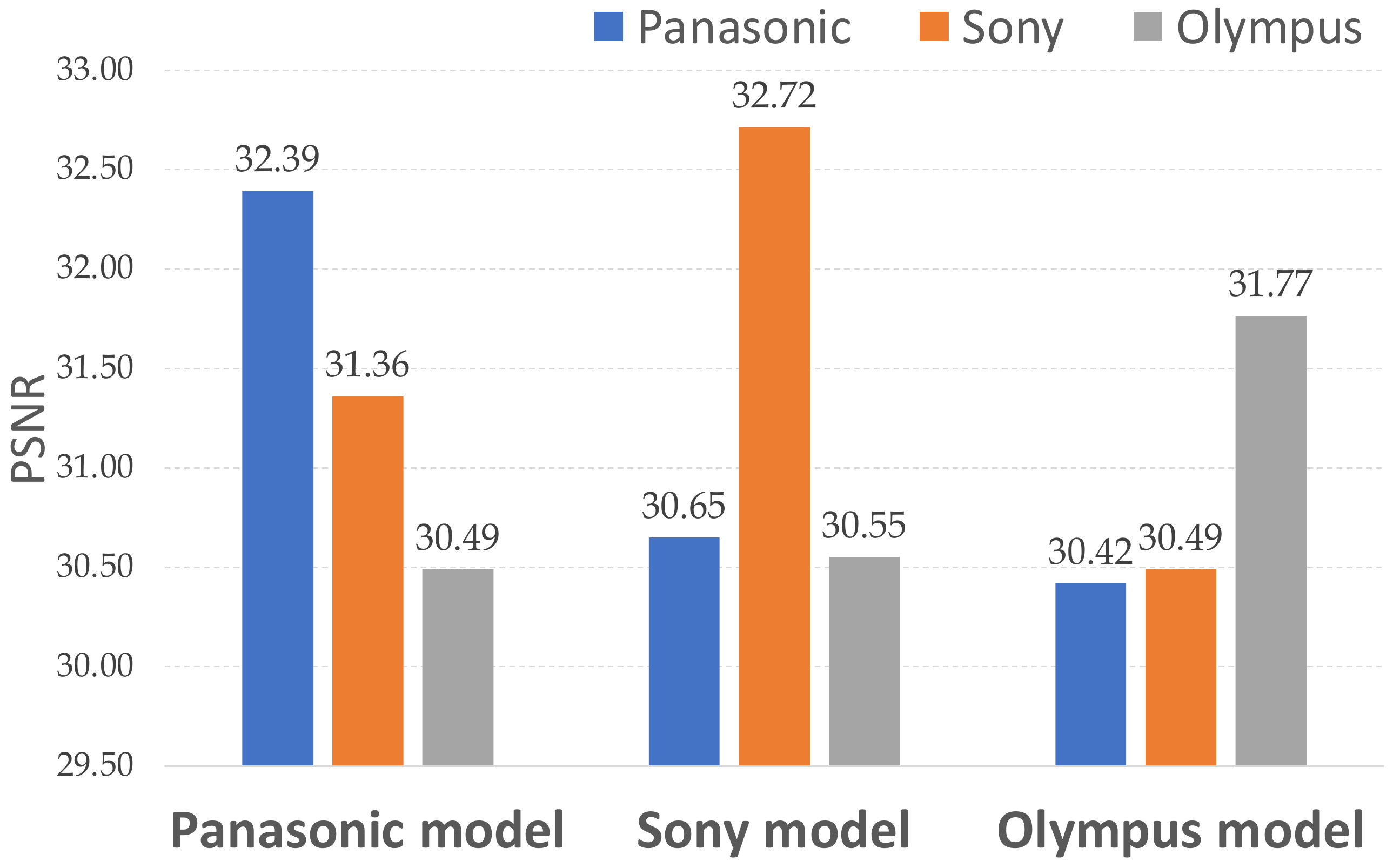}
  \vspace{-15pt}
  \caption{
  Cross-device evaluation of real SR models trained from images of individual cameras. We use models trained on the data collected by different cameras to test the images taken by a specific camera. 
  }
  \vspace{-5pt}
  \label{fig:cross_device_testing}
\end{figure}
\begin{figure}[t]
  \centering
  \includegraphics[width=.48\textwidth]{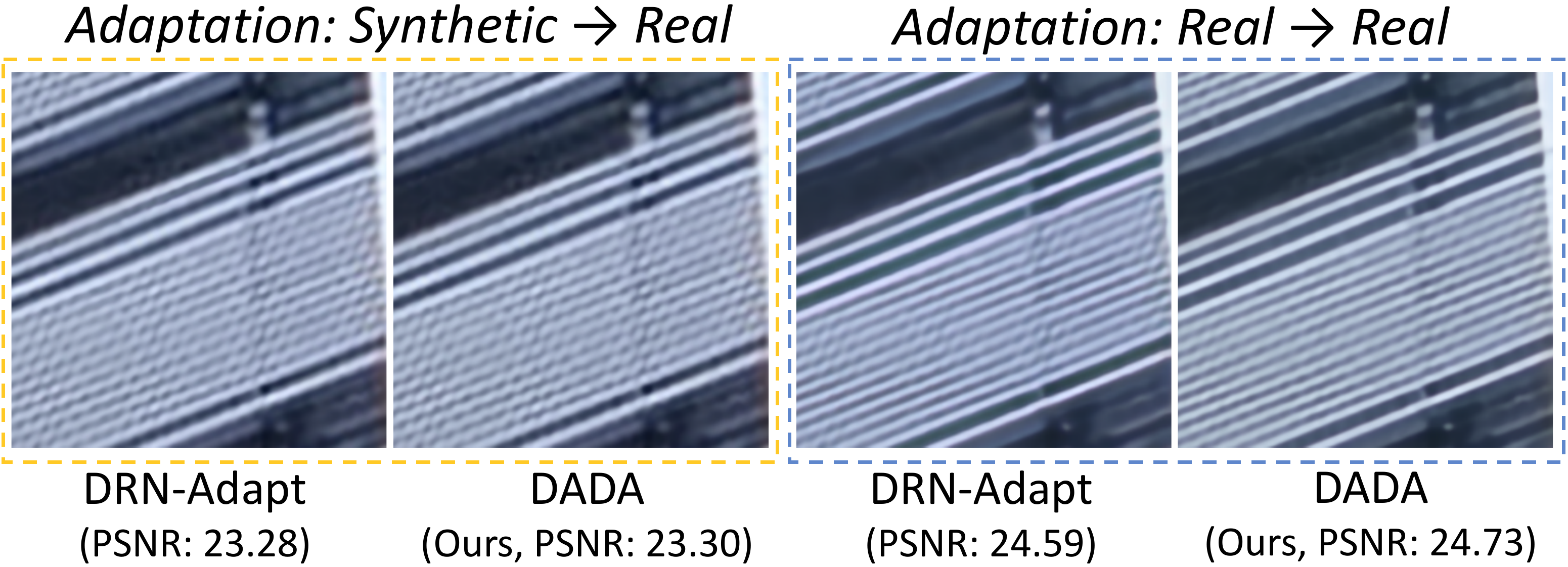}
  \vspace{-15pt}
  \caption{SR result comparison between synthetic to real adaptation and real to real adaptation (\textit{Sony $\rightarrow$ Panasonic}).}
  \vspace{-10pt}
  \label{fig:s2r_VS_r2r}
\end{figure}
In comparison with synthetic image degradation, real-world SR exhibits a crucial challenge, \emph{i.e.}, \emph{diverse degradation processes among devices}, due to different imaging processes across devices, especially across different camera types.
The \emph{cross-device domain gap} is demonstrated in Fig.~\ref{fig:diff}. It is observed that in cross-device real SR model evaluation (Sony$\rightarrow$Panasonic, or Panasonic$\rightarrow$Sony), it presents a distinct device-specific degradation gap: compared with that trained on Panasonic (Sony) images, an SR model trained only on Sony (Panasonic) images (\emph{i.e.}, \emph{Sony} (\emph{Panasonic}) model) super-resolve an LR image, captured by Panasonic (Sony) camera, into an SR image that has more blurry details or even distorted artifacts and larger differences from ground truth HR image. To explain this phenomenon,
degradation kernels for different (camera) degradations are analyzed in Fig.\ref{fig:kernel}. 
Real image degradation kernels are different across different cameras.
We regard device-specific degradation gap as \emph{domain gap} in this work and also empirically demonstrate its consequence of the performance degradation in cross-device/domain setting in Fig.~\ref{fig:cross_device_testing}.
However, this domain gap is ubiquitous for many realistic applications, \emph{e.g.}, image/video enhancement for all kinds of phones or GoPro cameras, and classic old movie restoration. Usually, it is extremely labor-intensive and difficult to collect paired data for each camera, and even is impossible to obtain paired data, \emph{e.g.}, classic old movies.

To mitigate this issue, we are the first to explore Unsupervised Domain Adaptation (UDA) for Real-World Image Super-Resolution across Devices.
Under this setting, given paired real LR-HR images captured by one camera (\emph{source camera/domain}), the goal is to adapt the model to the \emph{target domain} that has only LR images captured by another camera (\emph{Real} $\rightarrow$ \emph{Real} adaptation).
This is more rational than conventional UDA SR from source domain with paired synthesized LR-HR images to target domain with real images (\emph{Synthetic} $\rightarrow$ \emph{Real} adaptation).
As shown in Fig.~\ref{fig:kernel}, synthetic degradation (\emph{e.g.}, widely-used bicubic downsampling) is engaged with simple kernels; realistic degradation is heterogeneous and more complex.
The complexity of real kernels \textcolor{black}{brings} challenges to the \textit{Real $\rightarrow$ Real} adaptation task.
\textcolor{black}{Due to the significant distinction between synthetic and real degradation, it is certainly difficult to coordinate a source model of synthetic degradation to a target domain with realistic data, and has an inferior performance in the target domain.}
This is evidenced in Fig.~\ref{fig:s2r_VS_r2r}. Overall, our UDA SR across devices, namely \emph{Real} $\rightarrow$ \emph{Real} adaptation, is more practical for realistic applications.

In this paper, we propose a Dual ADversarial Adaptation model (DADA) to explore unsupervised domain adaptation for real-world image super-resolution across devices.
Rooted in the Component Divide-and-Conquer model (CDC)~\cite{CDC}, DADA has a source branch and a target branch in a symmetric architecture and each branch is essentially a cycle image reconstruction with an up-sampling module and a down-sampling module.
DADA employs the Domain-Invariant Attention (DIA) module to provide component guidance masks for up-sampling modules in two branches for each input.
Additionally, the dual framework of DADA facilitates an Inter-domain Adversarial Adaptation (InterAA) in \emph{one branch} for \emph{two LR input images} from two domains, and an Intra-domain Adversarial Adaptation (IntraAA) in \emph{two branches} for \emph{an LR input image}.
InterAA and IntraAA together improve the model transferability from \textcolor{black}{the} source domain to the target.

In summary, our main contributions are three-fold:
\begin{itemize}
\item
%
We are devoted to the early attempt to explore the cross-device domain gap in real-world image super-resolution. To mitigate this issue, a Dual ADversarial Adaptation model (DADA) is proposed for unsupervised domain adaptation from the source domain with paired real data to the target  with only real LR images.
\item We propose the Inter-domain adversarial Adaptation (InterAA) and the Intra-domain adversarial Adaptation (IntraAA) to train the model in a dual architecture for unsupervised \emph{Real} to \emph{Real} SR adaptation.
\item Extensive experiments on six \emph{Real} to \emph{Real} adaptation settings among three different camera \textcolor{black}{ domains demonstrate} the superiority of our DADA over traditional SR methods on real-world image super-resolution when adapting the model across different cameras.
\end{itemize}

\section{Related Work}
\subsection{Image Super-Resolution}
With a remarkable feature learning ability, Convolutional Neural Network (CNN) based methods have brought considerable improvements in the field of single image super-resolution compared to traditional methods.
SRCNN \cite{Dong2014LearningAD} is the first to employ the CNN network in the SR task, which is an end-to-end three-layer CNN network to learn the feature mapping from input LR images to HR images. Subsequently,
much deeper and more complex networks have been proposed, \emph{e.g.}, ESPCN \cite{shi2016real}, SRResNet \cite{Ledig2017PhotoRealisticSI}, EDSR \cite{Lim2017EnhancedDR}, SRDenseNet \cite{Tong2017ImageSU}, RCAN \cite{Zhang2018ImageSU} and ESRGAN \cite{Wang2018ESRGANES}, introducing structures such as dense connections, attention modules and non-local modules, and the SR performance is constantly improved.
However, due to the difficulty of collecting real LR-HR image pairs, those deep \textcolor{black}{learning-based} approaches synthetically down-sample HR images into their LR counterparts. With synthetic LR-HR pairs, it essentially casts SISR as a supervised learning problem.
However, simple downsampling methods cannot simulate the realistic degradation, causing performance degradation when directly applying the trained model in practice. 


\subsection{Real-World Image Super-Resolution}
To break the data bottleneck of synthetic degradation, real-world image super-resolution
\textcolor{black}{attracts increasing interest. }
A well-prepared real-world SR dataset is RealSR \cite{Cai2019TowardRS}, which has paired LR-HR images captured by zooming lens of DSLR cameras. Subsequently, DRealSR, as a more challenging large-scale real-world SR dataset, has been built with five DSLR cameras. Due to the vulnerability of deep neural networks, how adversarial perturbations affect SR models is explored for robust real-world super-resolution in the low-level vision~\cite{yue2021robust}.
However, existing real SR research treats images from different cameras equally without discrimination and the \textcolor{black}{cross-device domain gap}, brought by different realistic degradations derived from different cameras.
This is ignored in existing works, let alone conventional SR with synthetic image degradation (it is limited to uniform and simple down-sampling degradation kernels, \emph{e.g.}, bicubic down-sampling, for images no matter which camera they are captured by).

\subsection{Unsupervised Domain Adaptation in SR}
Most UDA researches are on high-level vision tasks, \emph{e.g.}, image classification, object detection and semantic segmentation \cite{ganin2016domain, tsai2018learning, zhu2019adapting, luo2019taking, saito2018maximum, chen2018domain, huang2020contextual}. Few are devoted to low-level vision, which is more challenging for UDA since it concentrates more on pixel-wise adaptation and is not easy just by simple feature alignment or sample distribution alignment like UDA in high-level vision.
Conventional UDA SR generally aims to leverage abundant paired synthetic LR-HR images (source domain) to transfer the model to super-resolve real images (target domain).
Motivated by CycleGAN \cite{Zhu2017UnpairedIT}, Yuan et al. \cite{CinCGAN} propose the CinCGAN model, to transfer real-world LR images to synthetic LR images.
Guo et al. \cite{DRN} propose a method that trains the synthetic paired data and unpaired real data together with cyclic constraints.
Wei et al. \cite{DASR} proposed a model named DASR, which uses unpaired real images to handle the real-world SR problem.
They use  domain-gap aware training and domain-distance weighted supervision strategies to narrow the domain gap between source data and target data.
However, these methods are not aware of the divergence of realistic image degradation between different cameras. We term this divergence as \textcolor{black}{cross-device domain gap} and propose a DADA model to address this \emph{Real} to \emph{Real} unsupervised domain adaptation.

\section{Methodology}
\begin{figure*}[t]
  \centering
  \includegraphics[width=0.95\textwidth]{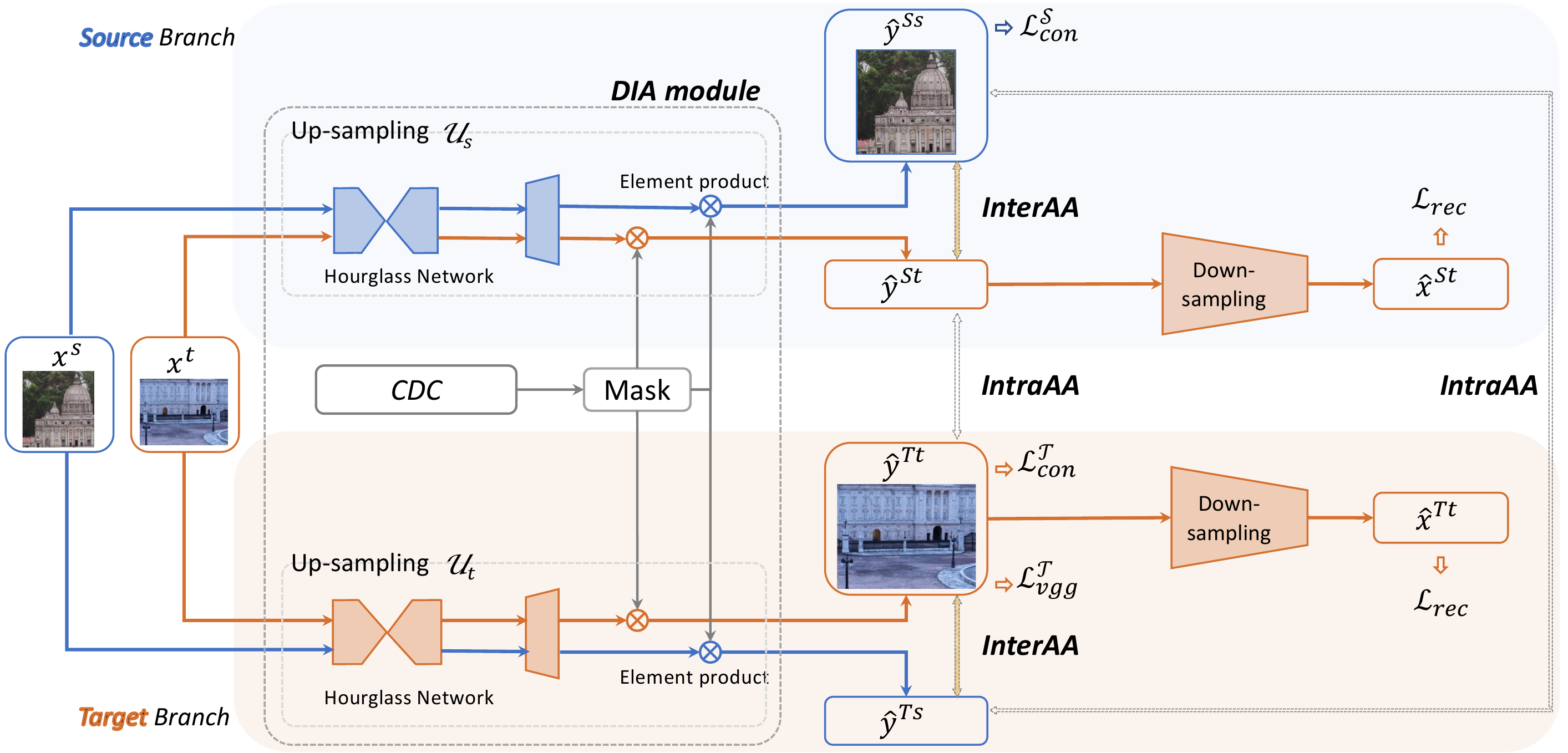}
  \vspace{-8pt}
  \caption{The proposed framework of our DADA. It includes a source branch and a target branch, both of which have an LR-HR-LR reconstruction structure. In their up-sampling networks,
  a Domain-Invariant Attention (DIA) module provides component attentive masks by a pre-trained CDC model. For InterAA, two LR images from different domains are fed into the up-sampling network for adversarial adaptation in one branch. For IntraAA, one LR image is respectively fed into up-sampling networks in source and target branches for the adversarial adaptation. For testing, the trained up-sampling network in target branch is utilized for inference.
  }
  \vspace{-10pt}
  \label{fig:framework}
\end{figure*}
We consider the real-world SR problem of unsupervised domain adaptation across devices, where real LR-HR image pairs from a camera are provided in the source domain, while only real LR images from another camera are accessible in the target domain. Noted that the device domain gap mentioned here stems from the imaging differences of two different models of cameras for the same scene\footnote{Different cameras might have different Image Signal Processors (ISPs) for image degradation.
Existing real SR datasets are collected by different cameras with different types/brands and do not include different cameras with the same ones. In this work, different cameras/devices specifically \textcolor{black}{indicate} different camera brands, namely different domains.}.
With source LR-HR real image pairs and target real LR images, we aim to train a UDA model for real SR in the target domain.
In this section, we will elaborate on the proposed DADA to explore the unsupervised domain adaptation problem for real-world image super-resolution across devices.

\subsection{Overview}
\vspace{-10pt}
(Close-set) %
UDA in high-level vision has an underlying prerequisite that the source domain and the target domain share the same semantic categories.
This is convenient to learn domain-invariant semantic features as the basis for model adaptation.
However, for UDA in real SR, it is intractable to figure out what is essentially the basis for adaptation.
To address this issue, on one hand, our DADA leverages the stability of extracting mid-level image components from low-level image pixels, to build a domain-invariant attention module, rather than learning domain-invariant features.
On the other hand, the proposed DADA model inherits the cycle-consistent reconstruction structure (\emph{LR}$\rightarrow$\emph{HR}$\rightarrow$\emph{LR}) for the target domain.
Considering the inaccessibility of HR images in \textcolor{black}{the} target domain,
DADA employs a training strategy with inter-domain and intra-domain adversarial adaptation in a dual architecture.
In Fig.~\ref{fig:framework}, our DADA consists of two symmetrical branches.
Each branch is a \emph{LR}$\rightarrow$\emph{HR}$\rightarrow$\emph{LR} reconstruction network, including an up-sampling module and a down-sampling module.
One branch is dominated by source data with paired LR-HR supervisions, named \emph{source branch}; the other is mainly responsible for the target domain, named \emph{target branch}.
Taking a source LR image $x^s$ and a target LR image $x^t$ as inputs, they are sent to the up-sampling modules in these two branches respectively, resulting in four SR outputs.
We use the \textit{Inter-domain Adversarial Adaptation} (InterAA) scheme to process SR outputs from different inputs of the same branch.
In contrast, the \textit{Intra-domain Adversarial Adaptation} (IntraAA) scheme processes SR outputs from \textcolor{black}{the} same inputs in different branches.
As stated in CDC~\cite{CDC}, image components (\emph{i.e.}, the flat, edges and corner points) are content-dependent, and thus they are domain-invariant and loosely influenced by different cameras. This motivates us to establish a domain-invariant attention module based on a pre-trained CDC. The two branches are trained collaboratively with inter-domain and intra-domain adversarial adaptation.

\subsection{Dual Adversarial Adaptation Model}
\vspace{-5pt}
For source domain, paired data $\{x_i^s, y_i^s\}_{i=1,...,M}$ are accessible, where $x_i^s$ is the LR image and $y_i^s$ is the corresponding HR image.
For \textcolor{black}{the} target domain, only LR images $\{x_j^t\}_{j=1,...,N}$ can be accessed and their HR counterparts $\{y_j^t\}$ are unknown.

DADA takes a source LR image $x_i^s$ and a target LR image $x_j^t$ as inputs.
Both LR images are fed into the source and target branches together, respectively.
In one branch, besides the \emph{LR}$\rightarrow$\emph{HR}$\rightarrow$\emph{LR} reconstruction process, $x_i^s$ and $x_j^t$ are re-solved into SR images by the same up-sampling module (a generator), which are further sent to a discriminator for source/target domain discrimination. 
We name this process \emph{Inter-domain Adversarial Adaptation} (InterAA).
Between two branches, a target (source) LR image goes through the \emph{LR}$\rightarrow$\emph{HR}$\rightarrow$\emph{LR} reconstruction in each branch, and the SR images generated by the upsampling modules of the two branches will be further sent to a branch discriminator to distinguish their branch source. We name this process \emph{Intra-domain Adversarial Adaptation} (IntraAA).



%
\begin{figure}[t]
  \centering
  \includegraphics[width=0.48\textwidth]{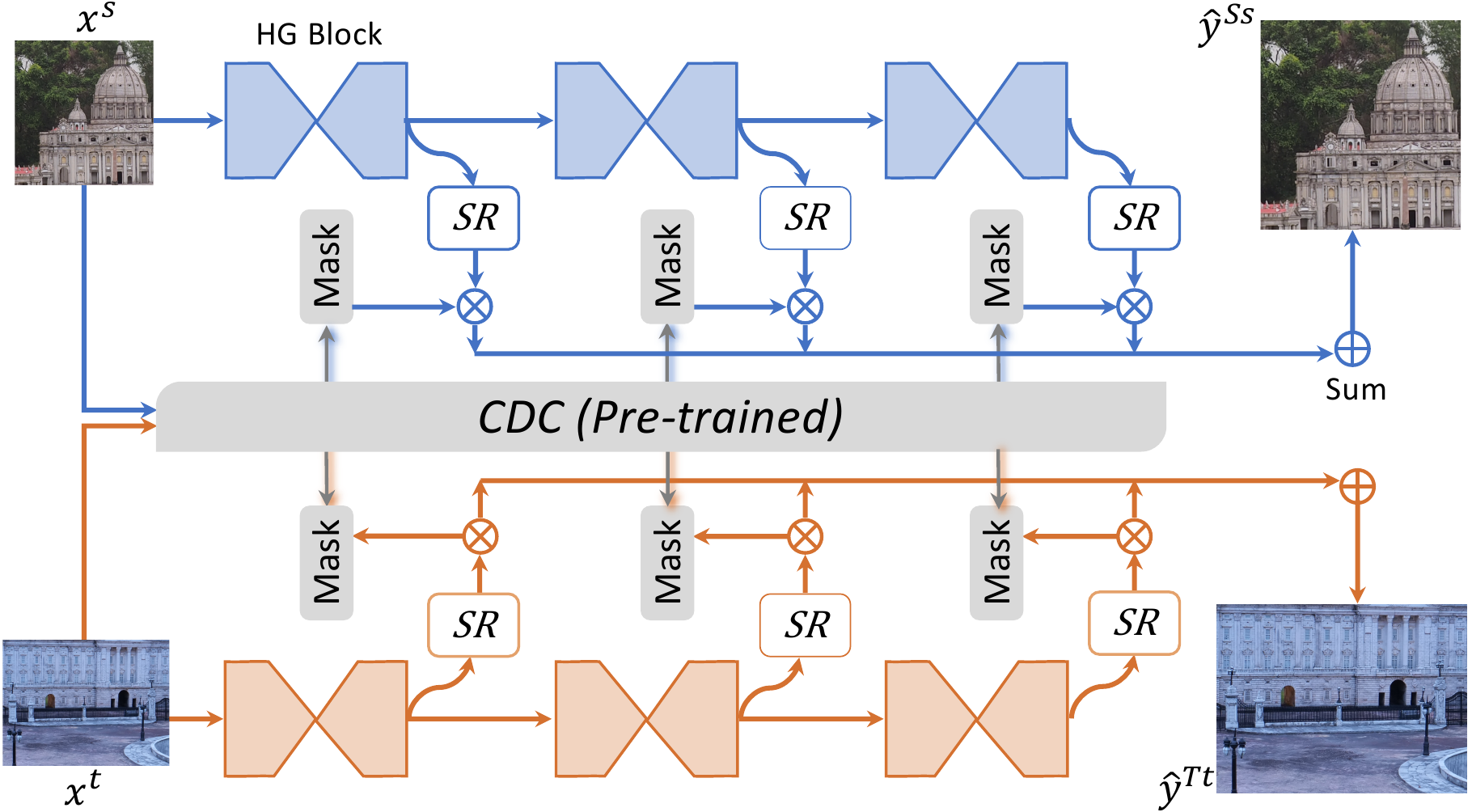}
  \vspace{-20pt}
  \caption{Domain-invariant attention module. For simplicity, we show its detailed network structure by taking two LR images as inputs in source and target branches, respectively.}
  \vspace{-12pt}
  \label{fig:module}
\end{figure}
%
\noindent
\textbf{Domain-Invariant Attention Module.}
We denote the up-sampling modules as $\mathcal{U}_s$ for the source branch and $\mathcal{U}_t$ for the target branch.
They follow the same neural network architecture as CDC~\cite{CDC}.
With the hourglass network as the backbone,  CDC builds three component-attentive blocks to address the issue of model overfitting to the simple image regions or contents for reconstruction.
The three image components are the flat, edges and corner points, respectively.
Considering those components are relatively invariant to the camera hardwares, it is believed that they are invariant for cross-device SR.
This is regarded as the root for real to real adaptation.
Accordingly, DADA utilizes a CDC model pre-trained on source pairs to provide domain-invariant attention masks (\emph{i.e.}, component masks in CDC) for LR input images.
Namely, the two up-sampling modules in each branch share the same parameter weights to generate component masks, as shown in Fig.~\ref{fig:module}.
Like CDC, those masks weigh three intermediate SR results, respectively, and then sum weighted results to generate the final SR result.

\noindent
\textbf{Inter-domain Adversarial Adaptation.}
In each branch, LR images from source domain and target domain are handled by the same up-sampling module.
There are GT HR images for source inputs, thus we can impose content supervision on the source SR images.
\textcolor{black}{
The InterAA process is to align the target domain with the source domain.
}

In the source branch, the up-sampling module $\mathcal{U}_s$ forwards a source LR image $x_i^s$ and a target LR image $x_j^t$, respectively, and produces their corresponding super-resolution results $\hat{y}_i^{Ss}$ and $\hat{y}_j^{St}$.
For those two SR images, \emph{i.e.}, the source SR result and the target SR result in the source branch, the discriminator $\mathcal{D}^{\textrm{inter}}_s$ is employed to distinguish which domain they come from.
By imposing an adversarial loss between $\hat{y}_i^{Ss}$ and $\hat{y}_j^{St}$, we force the network to produce results close to the source domain for the target input.
In this way, domain alignment is achieved in an adversarial manner, thereby improving the model's ability to process target data.
It is worth noting that since the two LR inputs from the two domains are very different, \emph{e.g.}, content and color, which increases the difficulty of adversarial training between them, we let the discriminator distinguish their SR images on the $Y$ channel.
This can avoid the bias to the image content or style.

Symmetrically, in the target branch, both the source LR image and the target LR image are up-sampled by $\mathcal{U}_t$, generating two SR results $\hat{y}^{Ts}$ and $\hat{y}^{Tt}$, which are then discriminated by the discriminator $\mathcal{D}^{\textrm{inter}}_t$.
There are two differences here from the source branch:
1) to avoid strong and dominant supervision from the source domain for $\mathcal{U}_t$ in the target branch, there is no supervision for the source SR $\hat{y}_i^{Ts}$.
2) Since source supervision is not used in the target branch, to stabilize the training of $\mathcal{U}_t$, we impose supervision on target SR images.
Due to the lack of target HR images, we use a pseudo-label for the target SR $\hat{y}_j^{Tt}$.
Specifically, we use the SR result $\hat{y}_j^{St}$ of the target input in the source branch as the label for $\hat{y}_j^{Tt}$.

Thus, in \textcolor{black}{the} source branch, the InterAA adversarial loss for the generator $\mathcal{U}_s$ and the discriminator $\mathcal{D}^{\textrm{inter}}_s$ is defined respectively as follows,
\vspace{-3mm}
\begin{equation}
  \begin{split}
    \label{equ:s_interAA_loss_g}
      \ & \mathcal{L}_{\textrm{inter}}^{\mathcal{S, G}} =  \mathbb{E}_{x_i^s} [log(1 - \mathcal{D}^{\textrm{inter}}_s(\mathcal{U}_s(x_i^s)))],
  \end{split}
\end{equation}
\vspace{-3mm}
\begin{equation}
  \begin{split}
    \label{equ:s_interAA_loss_d}
      \mathcal{L}_{\textrm{inter}}^{\mathcal{S, D}} =\ & \mathbb{E}_{x_j^t}[log( 1 - \mathcal{D}^{\textrm{inter}}_s( \mathcal{U}_s(x_j^t) ))]+ \\ &
      \mathbb{E}_{x_i^s} [log( \mathcal{D}^{\textrm{inter}}_s( \mathcal{U}_s(x_i^s) ))].
  \end{split}
\end{equation}

Symmetrically, in the target branch, the adversarial loss of InterAA for the generator $\mathcal{U}_t$ and the discriminator $\mathcal{D}^{\textrm{inter}}_t$ is respectively defined as follows,
\vspace{-3mm}
\begin{equation}
  \begin{split}
    \label{equ:t_interAA_loss_g}
      \ & \mathcal{L}_{\textrm{inter}}^{\mathcal{T, G}} = \mathbb{E}_{x_j^t} [log(1 - \mathcal{D}^{\textrm{inter}}_t(\mathcal{U}_t(x_j^t)))],  \end{split}
\end{equation}
\vspace{-3mm}
\begin{equation}
  \begin{split}
    \label{equ:t_interAA_loss_d}
      \mathcal{L}_{\textrm{inter}}^{\mathcal{T, D}} = \ &
      \mathbb{E}_{x_i^s} [log(1 - \mathcal{D}^{\textrm{inter}}_t( \mathcal{U}_t(x_i^s) ))] +
      \\ &  \mathbb{E}_{x_j^t} [log(\mathcal{D}^{\textrm{inter}}_t( \mathcal{U}_t(x_j^t) ))].
  \end{split}
\end{equation}


\noindent
\textbf{Intra-domain Adversarial Adaptation.}
InterAA adapts the model in the source or target branch with different LR images from two domains as inputs.
Oppositely, DADA utilizes intra-domain adversarial adaptation between source and target branches by taking the same LR images as inputs.
Specifically, taking the source LR image $x_i^s$ as an input of two branches, i.e., the two branches employ $\mathcal{U}_s$ and $\mathcal{U}_t$ to super-resolve it into SR images, $\hat{y}_i^{Ss}$ and $\hat{y}_i^{Ts}$, respectively.
The discriminator $\mathcal{D}^{\textrm{intra}}_{s}$ is used to identify which branch $\hat{y}_i^{Ss}$ and $\hat{y}_i^{Ts}$ are generated from.
Similarly, the target LR image will be sent to the two up-sampling modules in the two branches to obtain two SR counterparts, and then be distinguished by the discriminator $\mathcal{D}^{\textrm{intra}}_t$.
%
Namely, the two up-sampling modules, $\mathcal{U}_s$ and $\mathcal{U}_t$ cooperatively fool $\mathcal{D}^{\textrm{intra}}_{t}$.
By adopting IntraAA, we force the two up-sampling modules in the two branches to produce results that are close to each other, even though they are under different supervision, namely the source GT HR supervision and the target pseudo SR supervision.
In this way, we impose indirect and adversarial supervision on \textcolor{black}{the} target branch.
%

Accordingly, for a source LR image, the adversarial loss of IntraAA for the generator $\mathcal{U}_s$ and the discriminator $\mathcal{D}^{\textrm{intra}}_s$ is respectively defined as
\vspace{-5pt}
\begin{equation}
  \begin{split}
    \label{equ:s_intraAA_loss_g}
      \ & \mathcal{L}_{\textrm{intra}}^{\mathcal{S, G}} = \mathbb{E}_{x_i^s} [log(1 - \mathcal{D}^{\textrm{intra}}_s(\mathcal{U}_s(x_i^s)))],
  \end{split}
\end{equation}
\begin{equation}
  \begin{split}
    \label{equ:s_intraAA_loss}
      \mathcal{L}_{\textrm{intra}}^{\mathcal{S, D}} = \ &  \mathbb{E}_{x_i^s} [log(1 -  \mathcal{D}^{\textrm{intra}}_s( \mathcal{U}_t(x_i^s) ))] +
      \\ &  \mathbb{E}_{x_i^s}[log( \mathcal{D}_s^{\textrm{intra}}( \mathcal{U}_s(x_i^s) ))].
  \end{split}
\end{equation}
Similarly, for a target LR image, the adversarial loss of IntraAA for the generator $\mathcal{U}_t$ and the discriminator $\mathcal{D}^{\textrm{intra}}_t$ is respectively defined as
\vspace{-5pt}
\begin{equation}
  \begin{split}
    \label{equ:s_intraAA_loss_g}
      \ & \mathcal{L}_{\textrm{intra}}^{\mathcal{T, G}} = \mathbb{E}_{x_j^t} [log(1 - \mathcal{D}^{\textrm{intra}}_t(\mathcal{U}_t(x_j^t)))],
  \end{split}
\end{equation}
\begin{equation}
  \begin{split}
    \label{equ:s_intraAA_loss}
      \mathcal{L}_{\textrm{intra}}^{\mathcal{T, D}} = \ &  \mathbb{E}_{x_j^t} [log(1 - \mathcal{D}^{\textrm{intra}}_t( \mathcal{U}_s(x_j^t) ))] +
      \\ &  \mathbb{E}_{x_j^t}[log(\mathcal{D}_t^{\textrm{intra}}( \mathcal{U}_t(x_j^t) ))].
  \end{split}
\end{equation}

\subsection{Training Objective}
\vspace{-5pt}
Our training objective loss function $\mathcal{L}$ includes the
reconstruction loss $\mathcal{L}_{\textrm{rec}}$,
source content loss $\mathcal{L}_{con}^\mathcal{S}$,
target content loss $\mathcal{L}_{con}^\mathcal{T}$,
target VGG loss $\mathcal{L}^{\mathcal{T}}_{\textrm{vgg}}$
and the aforementioned adversarial loss of InterAA and IntraAA.
\begin{equation}
  \begin{split}
    \label{equ:total_loss}
    \mathcal{L} = \ &
                      \mathcal{L}_{con}^\mathcal{S} +
                      \mathcal{L}_{con}^\mathcal{T} +
                      \alpha \mathcal{L}_{\textrm{rec}} +
                      \beta \mathcal{L}^{\mathcal{T}}_{\textrm{vgg}} +
                 {\lambda}_1 \mathcal{L}_{\textrm{inter}}^{\mathcal{S,G}} + \\
                &
                \lambda_2 \mathcal{L}_{\textrm{inter}}^{\mathcal{T,G}} +
                \lambda_3 \mathcal{L}_{\textrm{intra}}^{\mathcal{S,G}}  +
                \lambda_4 \mathcal{L}_{\textrm{intra}}^{\mathcal{T,G}},
  \end{split}
\end{equation}
where hyper-parameters $\alpha$, $\beta$ and
$\lambda_1\sim\lambda_4$ are weight scalars.
%
\textbf{Reconstruction loss $\mathcal L_{rec}$:}
Each branch follows a cycle reconstruction framework (\emph{LR}$\rightarrow$\emph{HR}$\rightarrow$\emph{LR}).
For each input, we compute the $L_1$ reconstruction loss in each branch.
%
\textbf{Content loss $\mathcal L_{con}$:}
The Gradient Weighted (GW) loss~\cite{CDC} is utilized to compute the pixel-wise content loss. In source branch, $\mathcal{L}_{con}^\mathcal{S} = \mathcal{L}_{GW}(\hat{y}^{Ss}_i, y_i^s)$. In target branch, $\mathcal{L}_{con}^\mathcal{T} = \mathcal{L}_{GW}(\hat{y}^{Tt}_j, \hat{y}^{St}_j)$.
%
%
%
%
%
%
\textbf{VGG loss $\mathcal{L}^{\mathcal{T}}_{\textrm{vgg}}$:}
In the target branch, to alleviate the negative effects brought by pseudo HRs, VGG loss is used to constrain the target up-sampling module $\mathcal{U}_t$.
VGG-19 \cite{VGG} is the feature extractor $\phi$ whose \textit{conv5\_3} features are used. $\mathcal{L}^{\mathcal{T}}_{\textrm{vgg}} = \mathbb{E}_{x_j^t} [ \| \phi (\hat{y}_j^{Tt}) - \phi (\Bar{y}_j^t) \Vert_1 ]$,
where $\Bar{y}_j^t = \mathcal{U}_{s_0}(x_j^t)$, where $\mathcal{U}_{s_0}$ is the CDC model pre-trained on source data.

\section{Experiments}
\begin{table*}[t]
  \centering
  \renewcommand\arraystretch{0.82}
  \fontsize{8.8pt}{\baselineskip}\selectfont
  \setlength\tabcolsep{9pt}

  \begin{tabular}{cccccccccc}
  \toprule\hline
  \multicolumn{1}{c|}{\multirow{+2}*{\textbf{Method}}}        & \multicolumn{3}{c|}{\textbf{\textcolor[RGB]{237,125,49}{Panasonic} $\rightarrow$ \textcolor[RGB]{0,180,0}{Sony}}}            & \multicolumn{3}{c|}{\textbf{\textcolor[RGB]{0,180,0}{Sony} $\rightarrow$ \textcolor[RGB]{237,125,49}{Panasonic}}}            & \multicolumn{3}{c}{\textbf{\textcolor[RGB]{46,117,182}{Olympus} $\rightarrow$   \textcolor[RGB]{237,125,49}{Panasonic}}} \\ \cline{2-10}
  \multicolumn{1}{c|}{\multirow{-2}*{\textbf{Method}}}                                        & PSNR $\uparrow$           & SSIM $\uparrow$           & \multicolumn{1}{c|}{LPIPS $\downarrow$}          & PSNR $\uparrow$           & SSIM $\uparrow$           & \multicolumn{1}{c|}{LPIPS $\downarrow$}          & PSNR $\uparrow$                & SSIM $\uparrow$                & LPIPS $\downarrow$              \\ \hline
  \rowcolor{gray!20}\multicolumn{10}{c}{\textit{Real$\rightarrow${}Real}}   \\ \hline
  \multicolumn{1}{c|}{\textit{Target Only}}                             & 32.72          & 0.854          & \multicolumn{1}{c|}{0.302}          & 32.39          & 0.846          & \multicolumn{1}{c|}{0.316}          & 32.39               & 0.846               & 0.316              \\
  \cdashline{1-10}[0.5pt/1.5pt]
  \multicolumn{1}{c|}{\textit{Source Only}}                             & 31.36          & 0.838          & \multicolumn{1}{c|}{{0.319}} & 30.65          & 0.820          & \multicolumn{1}{c|}{0.383}          & 30.42               & 0.818               & 0.372              \\
  \multicolumn{1}{c|}{CinCGAN  \cite{CinCGAN}} & 27.76          & 0.821          & \multicolumn{1}{c|}{0.391}          & 28.33          & 0.792          & \multicolumn{1}{c|}{0.410}          & 29.37               & 0.799               & 0.381              \\
  \multicolumn{1}{c|}{DASR  \cite{DASR}} & 30.08          & 0.777          & \multicolumn{1}{c|}{\textbf{0.269}}          & 30.45          & 0.772          & \multicolumn{1}{c|}{\textbf{0.316}}          & 30.06               & 0.785               & \textbf{0.272}              \\
  \multicolumn{1}{c|}{DRN-Adapt  \cite{DRN}}   & 31.85          & 0.845          & \multicolumn{1}{c|}{0.321}          & 30.96          & 0.821          & \multicolumn{1}{c|}{0.380}          & 30.80               & 0.822               & 0.356              \\
  \multicolumn{1}{c|}{DADA (Ours)}                             & \textbf{32.13} & \textbf{0.849} & \multicolumn{1}{c|}{0.327}          & \textbf{31.25} & \textbf{0.825} & \multicolumn{1}{c|}{{0.363}} & \textbf{31.27}      & \textbf{0.824}      & {0.348}     \\ \hline
  \rowcolor{gray!20}\multicolumn{10}{c}{\textit{Synthetic $\rightarrow$   Real}}                                                                                                                                                                                                                  \\ \hline
  \multicolumn{1}{c|}{\textit{Source Only}}                             & 31.39          & 0.829          & \multicolumn{1}{c|}{0.369}          & 30.43          & 0.807          & \multicolumn{1}{c|}{0.433}          & 30.42               & 0.808               & 0.437              \\
  \multicolumn{1}{c|}{CinCGAN  \cite{CinCGAN}} & 27.59          & 0.788          & \multicolumn{1}{c|}{0.405}          & 27.19          & 0.743          & \multicolumn{1}{c|}{{0.414}} & 28.38               & 0.739               & {0.422}     \\
  \multicolumn{1}{c|}{DASR  \cite{DASR}} & 29.95          & 0.764          & \multicolumn{1}{c|}{\textbf{0.298}}          & 29.79          & 0.749          & \multicolumn{1}{c|}{\textbf{0.339}} & 30.02               & 0.777               & \textbf{\textbf{0.293}}     \\
  \multicolumn{1}{c|}{DRN-Adapt  \cite{DRN}}   & 31.42          & 0.829          & \multicolumn{1}{c|}{{0.359}} & 30.47 & 0.808          & \multicolumn{1}{c|}{0.429}          & 30.45               & 0.808               & 0.433              \\
  \multicolumn{1}{c|}{DADA (Ours)}                             & \textbf{31.50} & \textbf{0.830} & \multicolumn{1}{c|}{0.369}          & \textbf{30.72}          & \textbf{0.809} & \multicolumn{1}{c|}{0.376}          & \textbf{30.74}      & \textbf{0.808}      & {0.362}              \\ \hline
  \multicolumn{1}{c|}{\multirow{2}{*}{\textbf{Method}}}        & \multicolumn{3}{c|}{\textbf{\textcolor[RGB]{237,125,49}{Panasonic}   $\rightarrow$ \textcolor[RGB]{46,117,182}{Olympus}}}        & \multicolumn{3}{c|}{\textbf{\textcolor[RGB]{0,180,0}{Sony} $\rightarrow$ \textcolor[RGB]{46,117,182}{Olympus}}}              & \multicolumn{3}{c}{\textbf{\textcolor[RGB]{46,117,182}{Olympus} $\rightarrow$ \textcolor[RGB]{0,180,0}{Sony}}}        \\ \cline{2-10}
   \multicolumn{1}{c|}{\multirow{-2}{*}{\textbf{Method}}}                                        & PSNR $\uparrow$           & SSIM $\uparrow$           & \multicolumn{1}{c|}{LPIPS $\downarrow$}          & PSNR $\uparrow$           & SSIM $\uparrow$           & \multicolumn{1}{c|}{LPIPS $\downarrow$}          & PSNR $\uparrow$                & SSIM $\uparrow$                & LPIPS $\downarrow$              \\ \hline
  \rowcolor{gray!15}\multicolumn{10}{c}{\textit{Real$\rightarrow${}Real}}                                                                                                                                                                                                                         \\ \hline
  \multicolumn{1}{c|}{\textit{Target Only}}                             & 31.77          & 0.833          & \multicolumn{1}{c|}{0.375}          & 31.77          & 0.833          & \multicolumn{1}{c|}{0.375}          & 32.72               & 0.854               & 0.302              \\
  \cdashline{1-10}[0.5pt/1.5pt]
  \multicolumn{1}{c|}{\textit{Source Only}}                    & 30.49          & 0.816          & \multicolumn{1}{c|}{0.439}          & 30.55          & 0.810          & \multicolumn{1}{c|}{0.457}          & 30.49               & 0.814               & 0.330              \\
  \multicolumn{1}{c|}{CinCGAN  \cite{CinCGAN}} & 28.85          & 0.791          & \multicolumn{1}{c|}{0.461}          & 30.17          & 0.814          & \multicolumn{1}{c|}{0.443}          & 30.05               & 0.823               & 0.365              \\
  \multicolumn{1}{c|}{DASR  \cite{DASR}} & 29.32          & 0.768          & \multicolumn{1}{c|}{\textbf{0.306}}          & 29.86          & 0.762          & \multicolumn{1}{c|}{\textbf{0.372}}          & 30.29               & 0.787               & \textbf{0.270}              \\
  \multicolumn{1}{c|}{DRN-Adapt  \cite{DRN}}   & 30.73          & 0.816          & \multicolumn{1}{c|}{0.431}          & 30.66          & 0.810          & \multicolumn{1}{c|}{0.459}          & 31.47               & 0.833               & {0.312}     \\
  \multicolumn{1}{c|}{DADA (Ours)}                             & \textbf{31.08} & \textbf{0.820} & \multicolumn{1}{c|}{{0.433}} & \textbf{31.08} & \textbf{0.817} & \multicolumn{1}{c|}{{0.438}} & \textbf{32.05}      & \textbf{0.843}      & 0.343              \\ \hline
  \rowcolor{gray!15} \multicolumn{10}{c}{\textit{Synthetic $\rightarrow$   Real}}                                                                                                                                                                                                                  \\ \hline
  \multicolumn{1}{c|}{\textit{Source Only}}                             & 30.08          & 0.799          & \multicolumn{1}{c|}{0.479}          & 30.08          & 0.799          & \multicolumn{1}{c|}{0.472}          & 31.41               & 0.828               & 0.371              \\
  \multicolumn{1}{c|}{CinCGAN  \cite{CinCGAN}} & 28.43          & 0.766          & \multicolumn{1}{c|}{0.407}          & 29.34          & 0.767          & \multicolumn{1}{c|}{0.451}          & 29.50               & 0.792               & 0.392              \\
  \multicolumn{1}{c|}{DASR  \cite{DASR}} & 28.30          & 0.752          & \multicolumn{1}{c|}{\textbf{0.375}}          & 29.51          & 0.755          & \multicolumn{1}{c|}{\textbf{0.402}}          & 29.40               & 0.737              & \textbf{0.327}             \\
  \multicolumn{1}{c|}{DRN-Adapt  \cite{DRN}}   & 30.11          & {0.799}          & \multicolumn{1}{c|}{0.475}          & 30.11          & 0.799          & \multicolumn{1}{c|}{0.473}          & 31.45               & 0.829               & {0.362}     \\
  \multicolumn{1}{c|}{DADA (Ours)}                             & \textbf{30.40} & \textbf{0.800} & \multicolumn{1}{c|}{0.403}          & \textbf{30.62} & \textbf{0.803} & \multicolumn{1}{c|}{{0.411}} & \textbf{31.52}      & \textbf{0.829}      & 0.355              \\ \hline\bottomrule
  \end{tabular}
  \vspace{-8pt}
  \caption{Performance evaluation for UDA real SR across devices. (\emph{Target Only} model are trained with GTs. Except \emph{Target Only}, the highest performance among UDA SR methods are highlighted in bold.)}
  \vspace{-10pt}
  \label{tab:compare}
\end{table*}

\subsection{Experimental Settings}
\vspace{-5pt}
\noindent
\textbf{Dataset.}
Our experiments have been conducted on DRealSR dataset \cite{CDC} for UDA of real-world SR across cameras.
DRealSR is the only real-world SR dataset that involves multiple cameras and has distinct indications about which camera each image was captured by. It is collected by five DSLR cameras, \emph{i.e.}, Panasonic, Sony, Olympus, Nikon and Canon.
In our experiments, image pairs from three cameras (Panasonic, Sony and Olympus) are chosen for training and testing.
For each camera, they are split into a training set and a testing set, respectively.
There are 197 image pairs of Panasonic, 145 pairs of Sony and 190 pairs of Olympus for training, and 20, 17 and 19 corresponding pairs for testing.
Similar to \cite{CDC}, in the training stage, LR images are cropped into patches with the size of $48*48$. Our experiments are conducted for $\times 4$ scaling factor.
\begin{figure*}[t]
  \centering
  \includegraphics[width=0.99\textwidth]{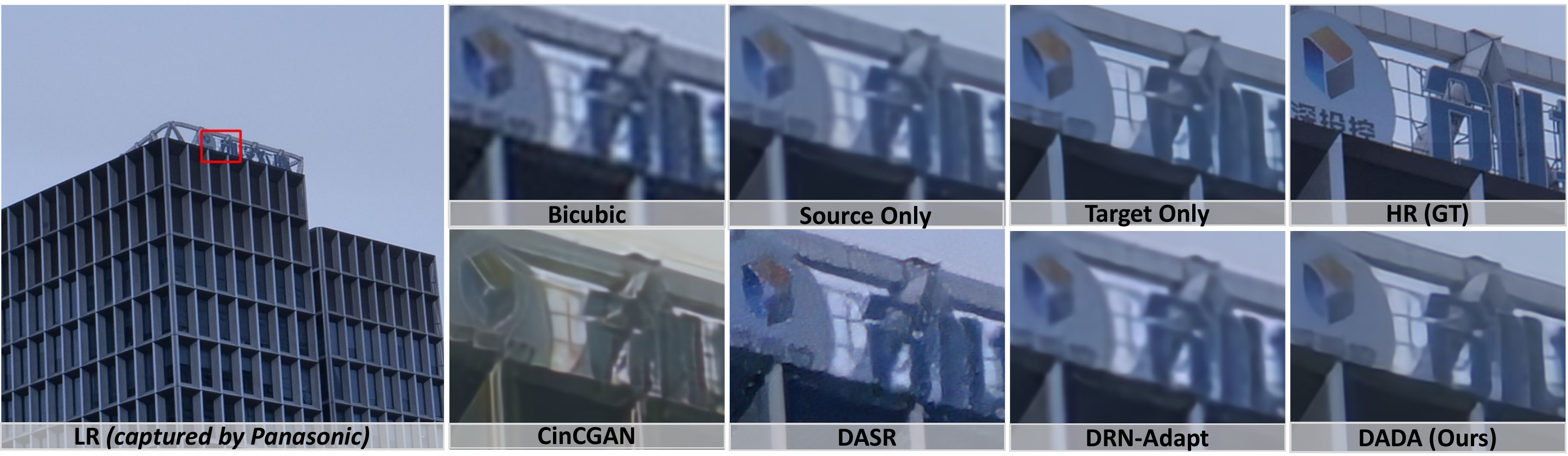}
  \vspace{-8pt}
  \caption{Comparison of UDA real SR results with state-of-the-art methods for \emph{Real} (Sony) to \emph{Real} (Panasonic) adaptation.}
  \vspace{-10pt}
  \label{fig:compare_s2p}
\end{figure*}

\noindent
\textbf{Implementation Details.}
We apply the Adam optimizer to train our model. The learning rate is $1e$-$4$. 
DADA takes a pair of a source image and a target image as an input, where source and target images are selected randomly. In each iterative step during training, four pairs are selected randomly as a batch.
%
Data augmentations includes random crop, random rotation and flip.
All the discriminators are PatchGAN \cite{Zhu2017UnpairedIT}.
The down-sampling module includes eight Residual blocks and two strided convolution layers \cite{DASR}.
For inference, the $\mathcal{U}_t$ network is employed to produce SR results.
$\lambda_1\sim\lambda_4$ are all set to 0.005 and $\alpha = 0.1, \beta = 0.01$.
Three commonly used metrics are adopted for evaluation, \emph{i.e.}, Peak Signal-to-Noise Ratio (PSNR), Structural Similarity Index (SSIM) \cite{Wang2004ImageQA} and LIPIS \cite{Zhang2018TheUE}. Following the similar setting to CDC\cite{CDC}, PSNR is computed on the $Y$ channel and SSIM is on RGB images.

\begin{figure*}[t]
  \centering
  \includegraphics[width=0.992\textwidth]{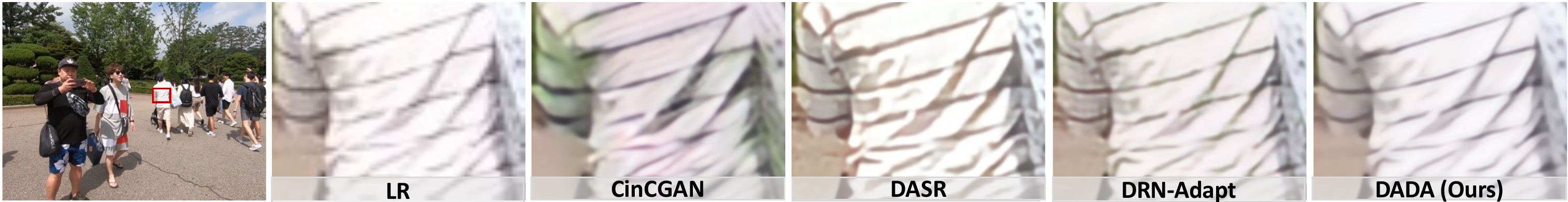}
  \vspace{-10pt}
  \caption{SR results of video images on the REDS dataset. For \emph{Real}$\rightarrow$\emph{Real} adaptation across devices, only qualitative visualizations of SR results are provided.
  Because only HR video images are realistic data, we take those HR video images in REDS as the target LR images to verify the proposed model.
  }
  \vspace{-15pt}
  \label{fig:compare_reds}
\end{figure*}

\vspace{-3pt}
\begin{figure}[tp]
\includegraphics[width=.48\textwidth]{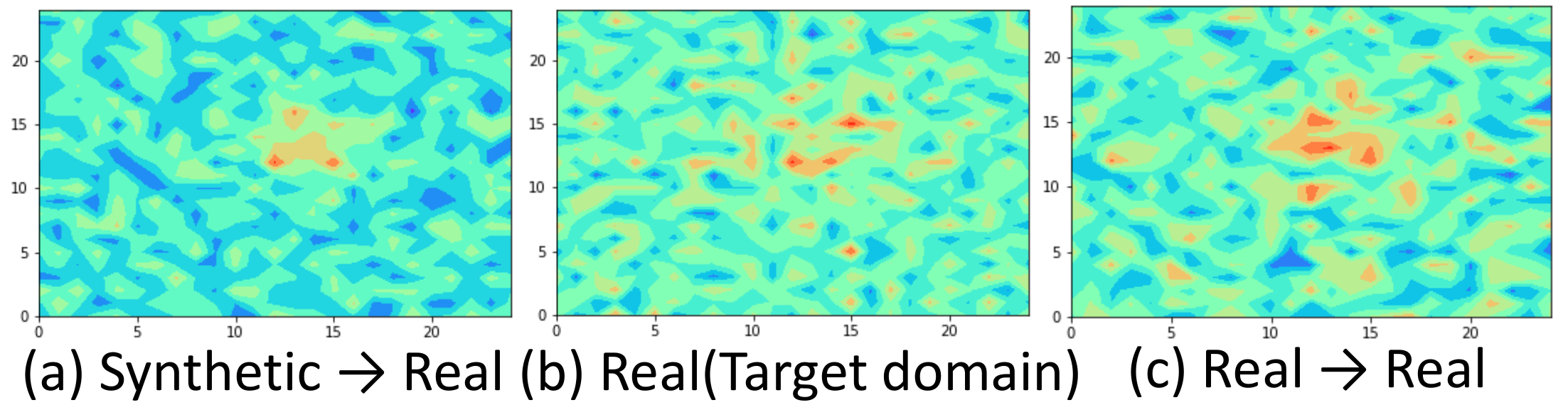}
\vspace{-18pt}
\caption{
Comparison of degradation kernels for generated SR images under different adaptations from \emph{Panasonic} \textcolor{black}{to} \emph{Sony}.
}
\vspace{-15pt}
\label{fig:sr_kernel}
\end{figure}

\subsection{Comparison with State-of-the-Art Methods}
\vspace{-5pt}
We are the first to explore the unsupervised domain adaptation problem between different devices w.r.t \textcolor{black}{the} super-resolution task.
We compare our method with existing \textit{Synthetic} to \textit{Real} UDA methods for SR,
including Cycle-in-Cycle Generative Adversarial Networks (CinCGAN) \cite{CinCGAN},  Domain-distance Aware Super-Resolution (DASR) \cite{DASR} and Dual Regression Adaptation Network (DRN-Adapt) \cite{DRN}.
For fair experimental comparisons, we implement them under the \textit{Real} to \textit{Real} adaptation setting by replacing the bicubic images with real LR images in \textcolor{black}{the} source domain.
The super-resolution network of CinCGAN is also the CDC model~\cite{CDC}.
%
%
Their comparison results are provided in Table \ref{tab:compare}. In this table, \textit{Source Only} is the model trained with paired source data without model adaptation.
\textit{Target Only} means the model trained with real paired data in the target domain.

\begin{table}[t]
  \small
  \centering
  \setlength\tabcolsep{4.5pt}
  \begin{tabular}{c|ccc|ccc}
  \hline
  Method                & InterAA      & IntraAA      & DIA          & PSNR  & SSIM  & LIPIS \\ \hline
  \multirow{4}{*}{DADA} &              & $\checkmark$ & $\checkmark$ & 30.76 & 0.812 & 0.472 \\
                        & $\checkmark$ &              & $\checkmark$ & 30.97 & 0.815 & 0.438 \\
                        & $\checkmark$ & $\checkmark$ &              & 31.03 & 0.816 & 0.445 \\
                        & $\checkmark$ & $\checkmark$ & $\checkmark$ & \textbf{31.08} & \textbf{0.817} & \textbf{0.438} \\ \hline
  \end{tabular}
  \vspace{-5pt}
  \caption{Ablation study.}
  \vspace{-20pt}
  \label{tab:ablation}
\end{table}

\vspace{2pt}
\noindent
\textbf{\textit{Real$\rightarrow$Real} \emph{Adaptation.}}
We conduct experimental comparisons under six settings of \textit{Real} to \textit{Real} adaptation among three cameras.
Compared with the baseline \emph{Source Only} model, our DADA achieves significant performance gains.
For example, in the \textit{Panasonic$\rightarrow$Sony} task, it improves the \emph{Source Only} model from 31.36dB to 32.13dB (PSNR).
In comparison with the state-of-the-art UDA methods, our method presents a superior performance and achieves best PSNR and SSIM on most of the six adaptation tasks.
For instance, in the \textit{Sony$\rightarrow$Olympus} adaptation task, our method outperforms CinCGAN by 0.91dB (PSNR), DASR by 1.22dB, and DRN-Adapt by 0.42dB, respectively.
Notably, in terms of the LPIPS metric, DASR achieves a rather high performance among all the methods including our DADA, however, its evaluations in terms of PSNR and SSIM are inferior in comparison with DRN-Adapt and our DADA. Though it indeed produces perceptually-clear SR images, it is of obvious noises and artifacts.
Particularly, Our DADA significantly outperforms DASR with PSNR gains of 0.80 dB (at least) and 2.05 dB (at most), and SSIM gains of 0.04 (at least) and 0.07 (at most).
In essence, the main reason to explain this phenomenon is that DASR employs adversarial training of the predicted SR images against the real-world HR images while DADA utilizes adversarial training between two predicted SR images. Thus, the former enforces DASR to produce SR images as similar (to HRs) as possible, but inevitably invites obvious noises and artifacts.
%
Besides, qualitatively, Fig.\ref{fig:compare_s2p} shows the visual comparison of different methods in the \textit{Sony $\rightarrow$ Panasonic} task.
It is observed that our DADA yields the visually better SR image than other comparison methods. For instance, the predicted SR image by \emph{Source Only} is blurred due to the device gap. 
The result of CinCGAN is sharp, but the color is distinctly different from the ground-truth HR image.
%
%
Instead, our DADA produces the clear result more close to the \textit{Target Only}.

\noindent
\textbf{\textit{Synthetic$\rightarrow$Real} \emph{Adaptation.}}
We also provide the evaluation results of \textit{Synthetic} to \textit{Real} adaptation in Table \ref{tab:compare}.
We use the synthetic pairs in the source dataset, where LR images are obtained by bicubic downsampling HR images.
It is observed that our DADA still achieves superior performance under six adaptation settings.
Moreover, in general, all the models have significantly high performance under \textit{Real} to \textit{Real} adaptation than those of under \textit{Synthetic} to \textit{Real} adaptation.
This demonstrates that the adaptation from \textit{Real} to \textit{Real} can achieve  higher image quality than that from \emph{Synthetic} to \emph{Real}.

\vspace{-3pt}
\subsection{Model Evaluation and Analysis}
\vspace{-5pt}
\noindent
\textbf{Ablation study.}
We conduct the model ablation studies in the \textit{Sony$\rightarrow$Olympus} task, as shown in Table \ref{tab:ablation}.
(1) \emph{DIA:} In the DIA module, for an input image, we share the attention mask on the source branch and the target branch.
Without DIA, we train the mask generator separately in each branch, and the result drops from 31.08dB to 31.03dB (PSNR).
(2) \emph{InterAA:} InterAA employs an SR model as the generator for source LR image and target LR image. It improves the model from 30.76dB to 31.08dB (PSNR).
(3) \emph{IntraAA:} IntraAA brings 0.11dB improvements.



\noindent
\textbf{Adaptation kernel analysis.}
In Fig.~\ref{fig:sr_kernel}, we employ USRNet to show degradation kernels for generated SR images under different adaptation. Fig.\ref{fig:sr_kernel}(b) is the GT real kernels for the target domain.
It is demonstrated that \emph{Synthetic} to \emph{Real} adaptation (Fig.~\ref{fig:sr_kernel}(a)) is limited by simple image degradation in the source domain and fails to fill the large domain gap cross devices. Instead, \emph{Real} to \emph{Real} adaptation (Fig.~\ref{fig:sr_kernel}(c)) presents a favorable transferring.

\noindent
\textbf{Evaluation for video camera.}
To fully verify our method, we additionally conduct an adaptation experiment to a video camera on REDS \cite{REDS}.
REDS is a video dataset proposed for video deblurring and super-resolution tasks, where each video is captured by the GoPro HERO6 Black camera.
Considering its LR frames are synthetic, we only use its original video frames as LR images for training.
We randomly select 22 videos for training and the rest are for testing.
In our work, temporal relationships of videos are not considered and \emph{Sony} in DRealSR is considered as the source domain.
In Fig.~\ref{fig:compare_reds}, we provide qualitative comparison results of \emph{Sony}$\rightarrow$\emph{REDS}.
It is observed that CinCGAN tends to produce SR results with wrong color and serious artifacts, and DASR also has chromatic aberration and distortion problems.
Our DADA has a clear and more natural SR result.
This presents a promising research interest, which would promote the practical applications of real-world SR to broad scenarios, \emph{e.g.}, video enhancement for handheld devices.

\vspace{-5pt}
\section{Conclusion}
\vspace{-2pt}

In this paper, we propose exploring the cross-device domain gap in the real-world super-resolution, facilitating an adaptation from a source domain with paired real LR-HR data to a target domain with only LR images.
To mitigate this issue, a Dual ADversarial Adaptation (DADA) model is proposed. It leverages the stability of extracting mid-level image components from images, to build \textcolor{black}{a} domain-invariant attention module, rather than learning domain-invariant features. Besides, considering the inaccessibility of HR images in target domains, our DADA employs a training strategy with inter-domain and intra-domain adversarial adaptation in a dual architecture.
Extensive experiments are conducted under six \emph{Real}$\rightarrow$\emph{Real} adaptation settings among three different cameras in the DRealSR dataset, demonstrating that the proposed DADA achieves
\textcolor{black}{superior}
performance compared with existing state-of-the-art approaches. Additionally, we also evaluate the proposed DADA to address the adaptation to \textcolor{black}{a} video camera, which presents a promising research topic to promote the wide applications of real-world super-resolution.

\noindent
\textbf{Broader impacts and limitations.} All the experiments are conducted on DRealSR, the only one dataset including multiple cameras with detailed device information, for cross-device real SR. The task might be more complex due to certain unknown issues when considering more different devices for image degradation. This task and the proposed method might need more analyses and evaluations.


{\small
\bibliographystyle{ieee_fullname}
\bibliography{main}
}

\end{document}